\documentstyle[hip-artc2,epsf]{article}

\recrevdate{June 6, 1997}                                               

\def\rightmark{Extracting the Equation of State through Hydrodynamical 
Analysis} \def\leftmark{B.R. Schlei}
 
\title{\vspace*{-1.0cm}
\hspace*{\fill}{\normalsize LA-UR-97-2162} \\[1.5ex]
Extracting the Equation of State of Nuclear Matter through Hydrodynamical 
Analysis}
\authors{
{\twerm B.R. Schlei$^{1,a}$ %
}\\[2.812mm]
{\normalsize
\hspace*{-8pt}$^1$ Theoretical Division, Los Alamos National Laboratory,\\ 
Los Alamos, NM 87545, USA
}}
 
\abstract{
Predominantly preliminary single and double inclusive momentum spectra of 158 $AGeV$
Pb+Pb collisions, recently measured by the NA44 and NA49 Collaborations,
are reproduced using the relativistic hydrodynamical model HYLANDER-C.
Two different equations of state, which both contain a phase transition
to a quark-gluon plasma, can be used to reproduce the (preliminary) data. 
The space-time geometries in the two calculations differ strongly. However,
the Bose-Einstein correlation functions of identical pion pairs do not show 
such a strong, but still a significant sensitivity to the effects of the 
equations of state.
} 
 
\begin{document}
 
\maketitle
 
\section{Introduction}
 
Fluid dynamics provides an intuitively simple description of heavy-ion 
collisions: two nuclei smash into each other and are rapidly thermalized and 
compressed; the resulting zone of very hot and dense nuclear matter (the 
so-called fireball) then expands and breaks up into bits of hadronic matter 
that ultimately reach the detectors. In a 158 $AGeV$ Pb+Pb reaction the 
velocity of the incoming projectile is 0.99998 $c$; since the sound speed in 
the ground state of nuclear matter is about $\sim$ 0.2 $c$, shock waves are 
formed. The density achieved in the center of the hot fireball is calculated 
to be as much as 20 times that of normal nuclear matter. Under such extreme 
conditions, a quark-gluon plasma (QGP) can be formed, which in our 
calculations can reach an appreciable fraction of the total matter 
\cite{bernd7},\cite{bernd8}.

Many experimental observables have been suggested with which one might 
ascertain whether a quark-gluon plasma (QGP) has been formed in relativistic 
heavy-ion collisions, or with which one can obtain some idea of the evolution 
of the experimentally generated fireballs. By measuring the correlations of 
identical particles (Bose-Einstein correlations (BEC)), one can get a measure 
of the sizes and lifetimes of the fireballs \cite{boal}. This is very similar 
to the concept behind the Hanbury-Brown/Twiss effect used to measure the size 
of stars \cite{GGLP}. 

\begin{figure}
\vspace*{-0.5cm}
\hspace*{-0.1cm}
                 \epsfxsize=12.5cm
		 \epsfbox{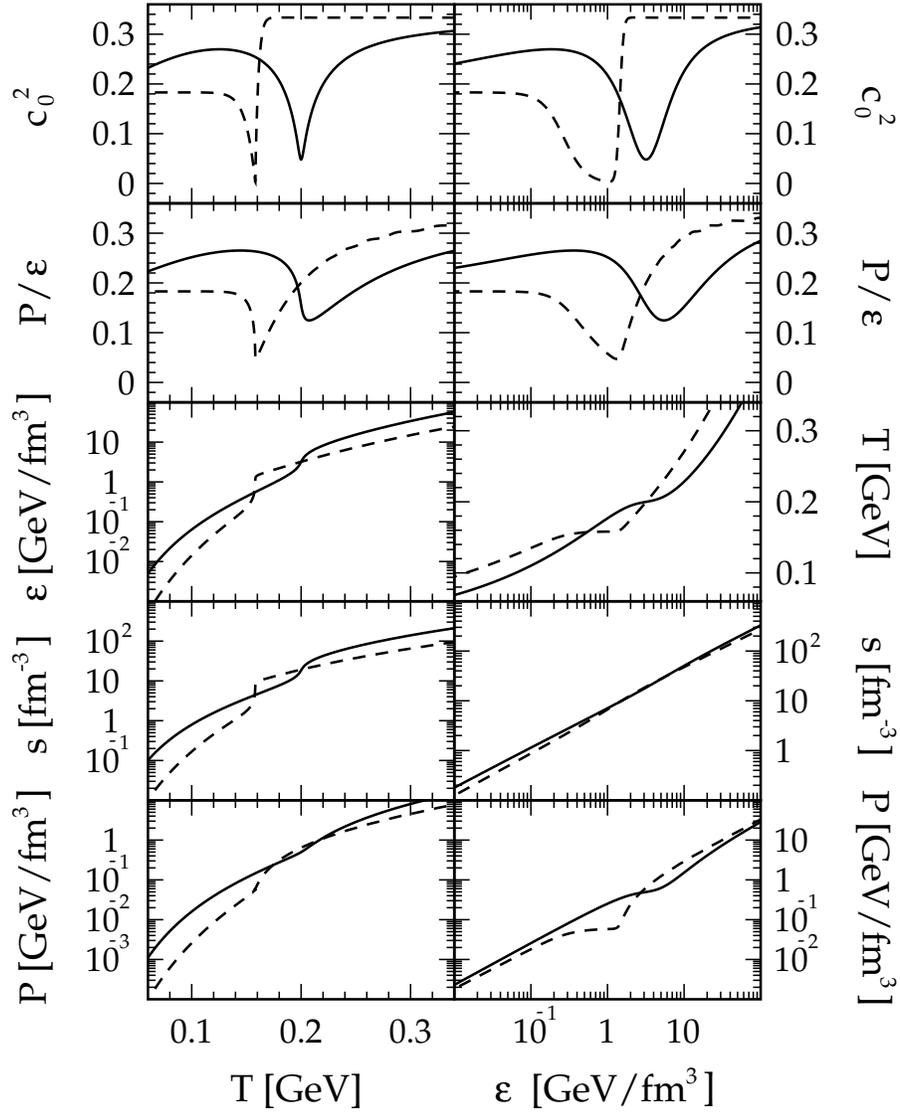}
\vspace*{0.0cm}
\caption[]{
Pressure, $P$, entropy density, $s$, energy density, $\epsilon$,
ratio of pressure and energy density, $P/\epsilon$, speed of sound,
$c_0^2$, and temperature, $T$, as functions of $T$ and/or $\epsilon$,
for the equations of state EOS-I (solid lines) and EOS-II (dashed lines), 
respectively (see text).
}
\label{fig01}
\end{figure}

It is the purpose of this paper to use in a hydrodynamical analysis two not 
too different equations of state (EOS) in the attempt to reproduce the predominantly
preliminary single inclusive momentum spectra of 158 $AGeV$ Pb+Pb collisions, 
recently measured by the NA44 \cite{NA44xu},\cite{NA44baerden} and NA49 
Collaborations \cite{NA49jones}, and to learn which details of the calculations are 
most sensitive to the effects of the EOS. We shall also learn about the initial 
conditions which one has to use while considering the chosen equations of state. 
After a comparison of the theoretical calculations
with the measurements, we exhibit the space-time geometries of the
real hadron sources in the two scenarios, and discuss their particular
features. Finally, we shall calculate inverse width parameters of Bose-Einstein 
correlation functions of identical pion pairs for the two different 
hydrodynamical solutions and compare the results to preliminary BEC data taken 
by the NA49 Collaboration \cite{NA49kadija}.
 
\section{The Model, the Equations of State, and the Initial Conditions}  

\subsection{The Model}  

Among the large number of models ({\it cf.} refs. \cite{bernd8},\cite{clare} 
and refs. therein) which apply relativistic hydrodynamics to relativistic 
heavy-ion collisions, we choose for the following to use the model$^b$ HYLANDER-C.

This model applies 3+1-dimensional relativistic one-fluid-dynamics, and
provides fully three-dimensional solutions of the hydrodynamical relativistic 
Euler-equations \cite{euler}. One has to specify an equation of state, initial 
distributions, e.g., in form of parameterizations with initial parameters 
({\it cf.}, refs. \cite{bernd8},\cite{jan}), and a break-up (freeze-out) condition 
in order to obtain an unambiguous solution from the hydrodynamical equations.
In the calculations we shall assume that hadronization occurs for all particle 
species at the same fixed freeze-out temperature $T_f$ = 139 $MeV$.

All model calculations are based on the assumption of thermal as well as on 
chemical equilibrium. In both types of spectra (single inclusive momentum
spectra and BEC, respectively) we shall include resonance decays. In particular, 
the momentum distributions are calculated in terms of the generalized Cooper-Frye 
formula (see ref. \cite{fred}), where explicitly a baryon and a strangeness 
chemical potential are taken into account \cite{jan}. The subsequent calculations 
of Bose-Einstein correlations are performed using the formalism outlined in ref. 
\cite{bernd3}. The hadron source is assumed to be fully chaotic; the influence of 
partial coherence \cite{bernd3} will not be considered here.

\subsection{The Equations of State}  

Any type of EOS can be considered when solving the relativistic Euler-equations. 
In Fig.~1 the two equations of state, which we are going to use in the following,
are plotted in many different representations.

The first equation of state, EOS-I, has been used in many recent calculations 
({\it cf.}, e.g. refs. \cite{bernd7},\cite{bernd8},\cite{jan},\cite{bernd3}), in 
the successful attempt to reproduce hadronic single inclusive momentum spectra and 
BEC, which have been measured in relativistic heavy-ion collision experiments.
In particular, EOS-I exhibits a phase transition to a quark-gluon plasma at 
a critical temperature $T_c$ = 200 $MeV$ with a critical energy density 
$\epsilon_c$ = 3.2 $GeV/fm^3$ ({\it cf.}, refs. 
\cite{redlich},\cite{phdudo},\cite{dipljan}).

The second equation of state, EOS-II, is also a lattice QCD based EOS which 
has recently become very popular in the field of relativistic heavy-ion 
physics ({\it cf.}, ref. \cite{hung}). This equation of state includes a phase 
transition to a quark-qluon plasma at $T_c$ = 160 $MeV$ with a critical 
energy density $\epsilon_c$ $\approx$ 1.5 $GeV/fm^3$.

Both EOS have no dependence on the baryon density. For instance, in Fig.~1 the 
plot of $P(\epsilon)/\epsilon$ emphasizes the existence of a minimum $P/\epsilon$
at $\epsilon$ = $\epsilon_c$ = 3.2 $GeV/fm^3$ ($\approx$ 1.5 $GeV/fm^3$)
for EOS-I (EOS-II), referred to as {\it the softest point} of the EOS.
It corresponds to the boundary between the generalized mixed phase and the 
QGP \cite{hung}. As it can be seen in Fig.~1, EOS-II yields a much softer 
equation of state than EOS-I.

\begin{figure}
\vspace*{0.0cm}
\hspace*{-0.2cm}
                 \epsfxsize=13.0cm
		 \epsfbox{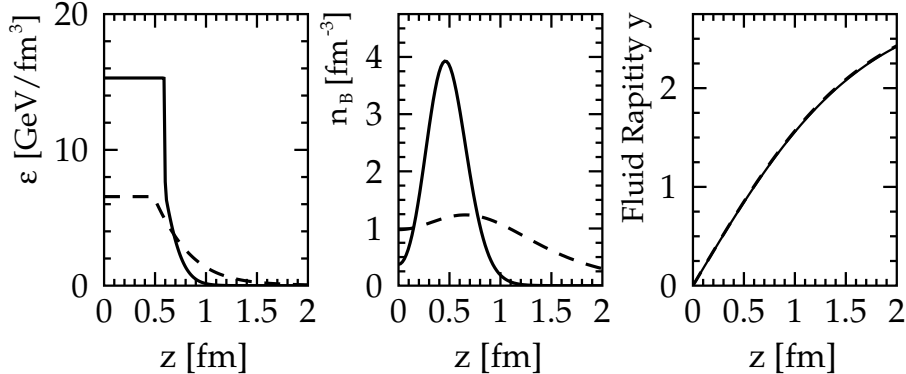}
\vspace*{-0.5cm}
\caption[]{Initial distributions of the energy density, $\epsilon$, the baryon 
density, $n_B$, and the fluid rapidity, $y$, plotted against the longitudinal 
coordinate $z$. The solid lines indicate the initial distributions for the use 
with EOS-I, whereas the dashed lines indicate the initial distributions for the 
use with EOS-II, respectively.
}
\vspace*{-0.5cm}
\label{fig02}
\end{figure}

\subsection{The Initial Conditions}

In the following it is assumed that due to an experimental uncertainty for the 
centrality of the collision, only 90$\%$ of the total available energy and the 
total baryon number have been observed. It is then possible to find initial 
distributions for the two here considered equations of state, such that one can 
reproduce the (preliminary) single inclusive momentum spectra of 158 $AGeV$ Pb+Pb 
collisions, which have been measured recently by the NA44 and NA49 Collaborations.

Fig.~2 shows the initial distributions for the energy density, $\epsilon(z)$, 
the baryon density, $n_B(z)$, and the fluid rapidity, $y(z)$, plotted against 
the longitudinal coordinate $z$.
We use here for the initial distributions the initial condition scenarios which 
have been extensively described in refs. \cite{bernd8} and \cite{jan}. In
particular, it is assumed that an initial transverse fluid velocity field is 
completely absent, and the initial longitudinal distributions for energy density, 
$\epsilon$, and baryon density, $n_B$, are smeared out with a Woods-Saxon 
parametrization in the transverse direction, $r_\perp$ ({\it cf.}, ref. 
\cite{bernd3}).
The choices for the ini\-tial parameters, which in each case provide the best fit 
results for the hadronic single inclusive momentum spectra of the two considered 
experiments, are in case of EOS-I (EOS-II): relative fraction of thermal energy
in the central fireball, $K_L$ = 0.55 (0.20), longitudinal extension of the
central fireball, $\Delta$ = 1.20 $fm$ (1.00 $fm$), rapidity at the edge of the
central fireball, $y_\Delta$ = 1.00 (0.85), rapidity at maximum of initial
baryon $y$ distribution, $y_m$ = 0.80 (1.50), and width of initial baryon $y$ 
distribution, $\sigma$ = 0.32 (1.00), respectively.

The maximum initial energy density is $\epsilon_\Delta$ = 15.3 $GeV/fm^3$ (6.55 
$GeV/fm^3$), and the maximum initial baryon density is $n_B^{max}$ = 3.93 
$fm^{-3}$ (1.24 $fm^{-3}$), for EOS-I (EOS-II), respectively. 71\% (30\%) of the
baryonic matter is initially located in the central fireball region. Hence, the 
use of EOS-I predicts a much larger stopping than the use of EOS-II.

It should be stressed here, that it is really neccessary to use such extremely
different initial conditions for EOS-I and EOS-II. If one just uses different 
equations of state without changing the initial conditions, which in fact are 
unknown for 158 $AGeV$ Pb+Pb collisions, one would have obtained single inclusive 
momentum spectra which differ by up to 20\% - 30\% when comparing the calculations
for EOS-I and EOS-II.

\section{Single Inclusive Momentum Spectra}

Figs. 3 - 6 show the results of the hydrodynamical calculations compared to the 
preliminary and final single inclusive momentum spectra of the 158 $AGeV$ Pb+Pb 
collisions, which have been measured by the NA44 \cite{NA44xu},\cite{NA44baerden} 
and NA49 Collaborations \cite{NA49jones}. All single inclusive momentum spectra 
have been evaluated in the nucleus-nucleus center of mass system ($y_{cm}$ = 2.91).

In case of EOS-I (EOS-II) the average value for the baryonic chemical potential 
is $\langle \mu_B \rangle$ = 324 $MeV$ (360 $MeV$), and the average value for 
the strangeness chemical potential is $\langle \mu_S \rangle$ = 55 $MeV$ 
(69 $MeV$), using $T_f$ = 139 $MeV$ for the freeze-out temperature. The energy 
density at freeze-out is $\epsilon_f$ = 0.292 $GeV/fm^3$ (0.126 $GeV/fm^3$).
Both calculations yield considerably good agreement with both experiments. 
However, the calculations for the much softer EOS-II already gives larger slopes 
in the transverse mass spectra compared to the calculations for EOS-I. The different 
slopes in the transverse mass spectra have their origin in the different transverse 
velocity fields at freeze-out. For EOS-I we obtain a maximum transverse velocity 
$v_\perp^{max}(I)$ = 0.46 $c$, whereas for EOS-II we get the much smaller value

\begin{figure}
\vspace*{-0.5cm}
\hspace*{-0.7cm}
                 \epsfxsize=14.0cm
		 \epsfbox{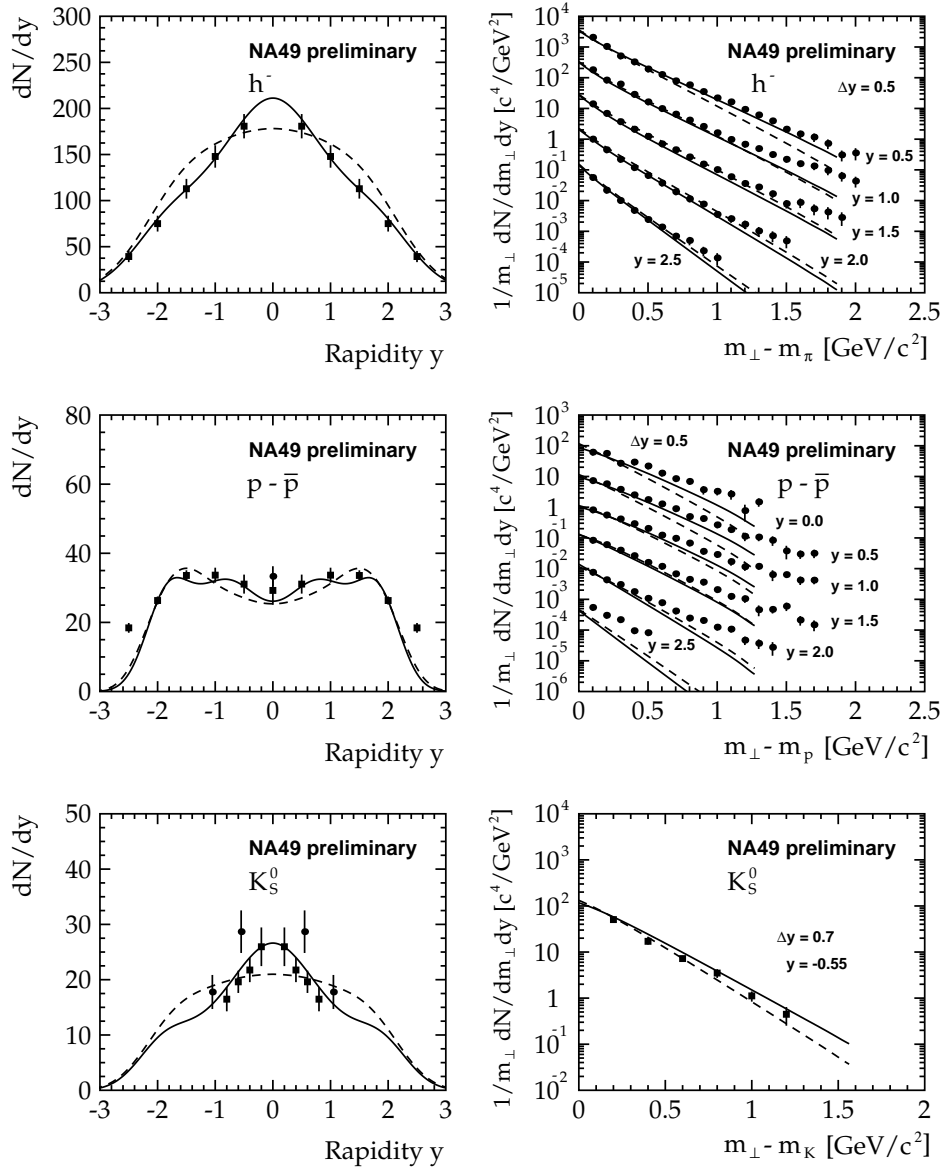}
\vspace*{-1.0cm}
\caption[]{Rapidity spectra, $dN/dy$, and transverse mass
spectra, $1/m_\perp dN/dm_\perp$, for negative hadrons, $h^-$ 
({\it i.e., $\pi^-$, $K^-$, $\bar{p}$}),
net protons (without contributions from $\Lambda^0$ decay), $p - \bar{p}$, 
and neutral kaons, $K^0_S$, respectively.
The solid lines indicate the results of the calculations when using equation 
of state EOS-I, whereas the dashed lines represent the results when using 
equation of state EOS-II. The shown preliminary data have been taken
by the NA49 Collaboration \cite{NA49jones}.
}
\label{fig03}
\end{figure}

\begin{figure}
\vspace*{-0.5cm}
\hspace*{-0.7cm}
                 \epsfxsize=14.0cm
		 \epsfbox{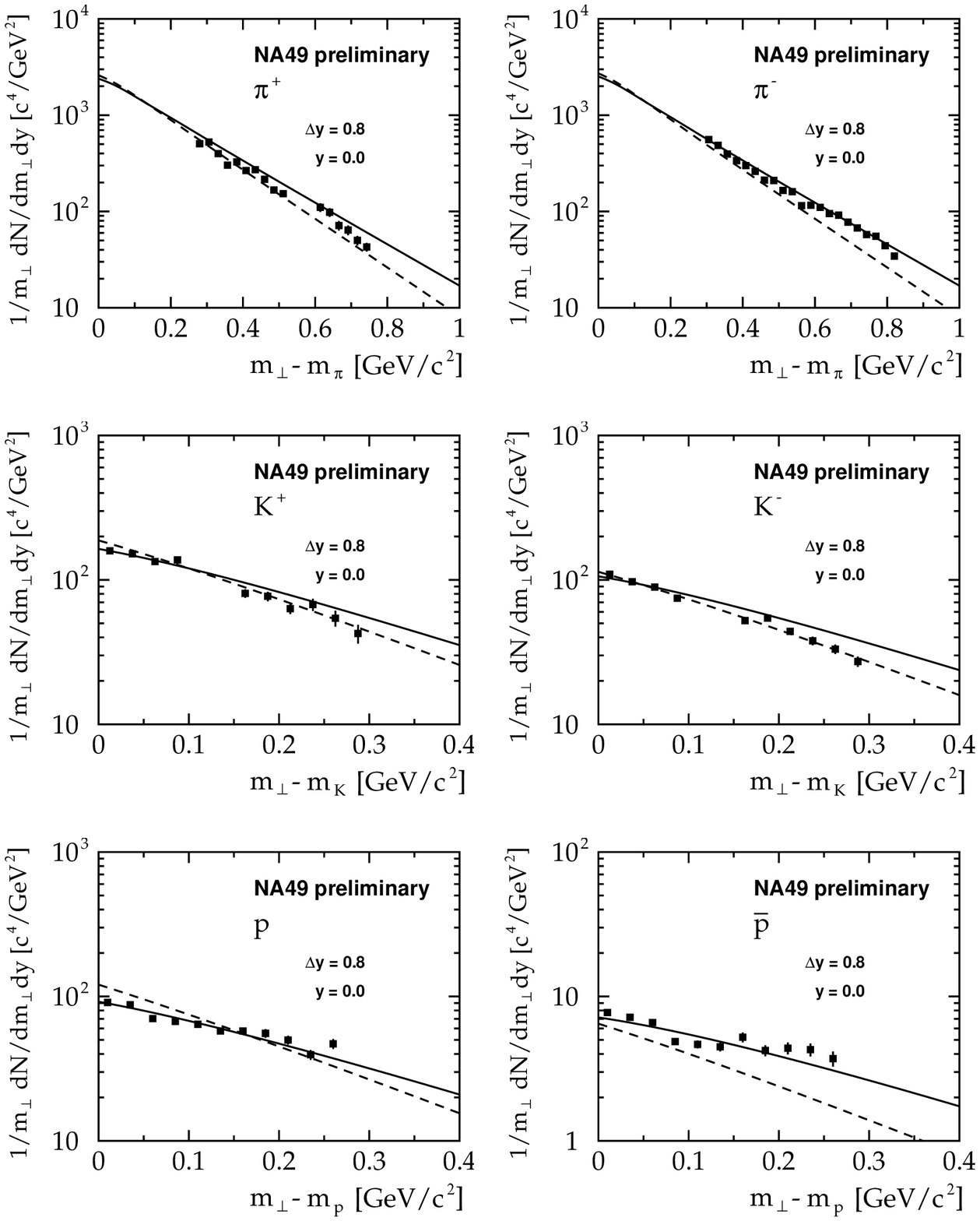}
\vspace*{-1.0cm}
\caption[]{Transverse mass spectra, $1/m_\perp dN/dm_\perp$, for pions, $\pi^+$
and $\pi^-$, kaons, $K^+$ and $K^-$, protons (without contributions from 
$\Lambda^0$ decay), $p$, and anti-protons, $\bar{p}$, respectively.
The solid lines indicate the results of the calculations when using equation 
of state EOS-I, whereas the dashed lines represent the results when using 
equation of state EOS-II. The shown preliminary data have been taken
by the NA49 Collaboration \cite{NA49jones}.
}
\label{fig04}
\end{figure}

\begin{figure}
\vspace*{-0.5cm}
\hspace*{-0.7cm}
                 \epsfxsize=14.0cm
		 \epsfbox{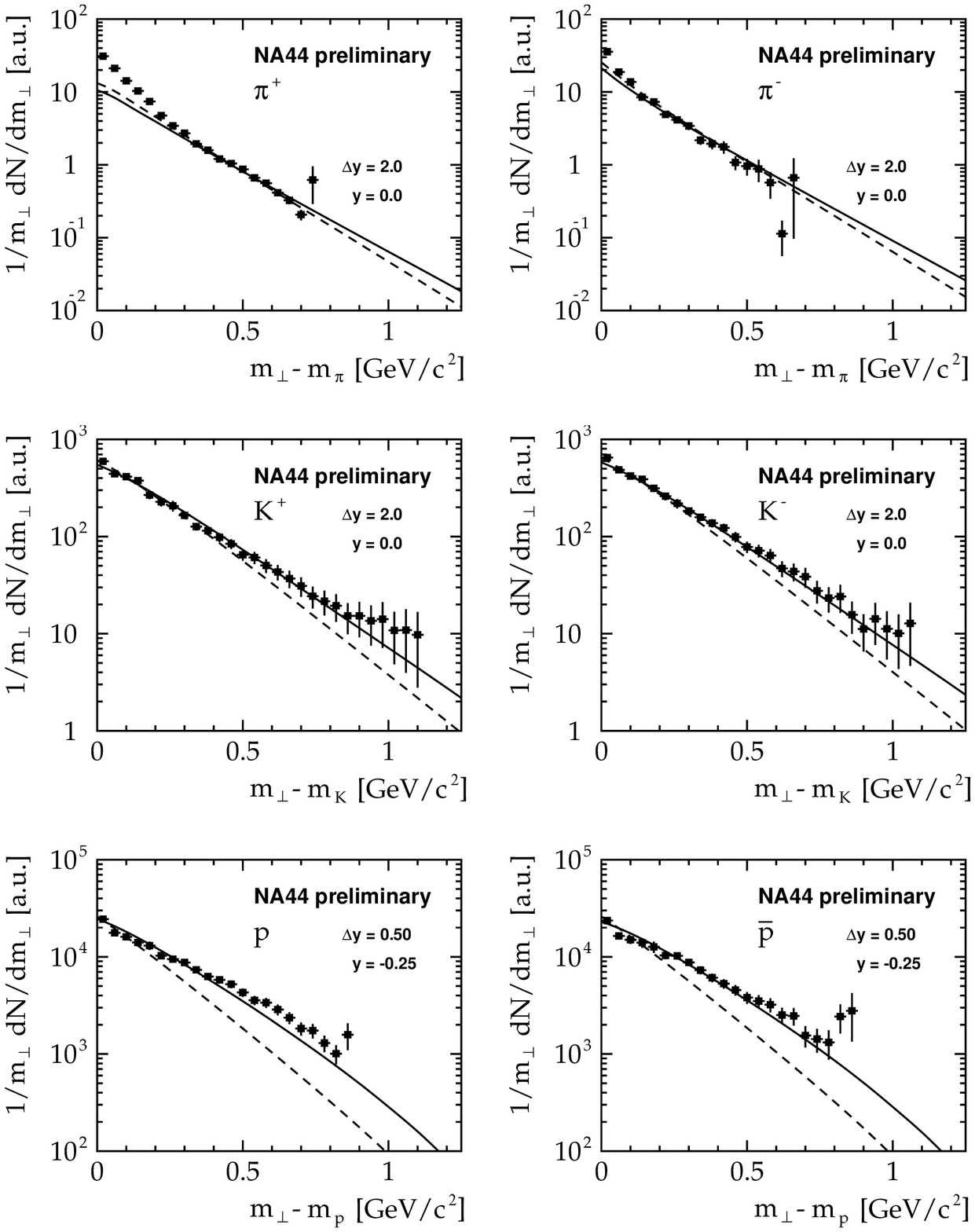}
\vspace*{-1.0cm}
\caption[]{Unnormalized transverse mass spectra, $1/m_\perp dN/dm_\perp$, 
for pions, $\pi^+$ and $\pi^-$, kaons, $K^+$ and $K^-$, protons 
(including contributions from $\Lambda^0$ decay), $p$, and anti-protons, 
$\bar{p}$, respectively.
The solid lines indicate the results of the calculations when using equation 
of state EOS-I, whereas the dashed lines represent the results when using 
equation of state EOS-II. The shown preliminary data have been taken
by the NA44 Collaboration \cite{NA44xu}.
}
\label{fig05}
\end{figure}

\begin{figure}
\vspace*{-0.5cm}
\hspace*{-0.7cm}
                 \epsfxsize=14.0cm
		 \epsfbox{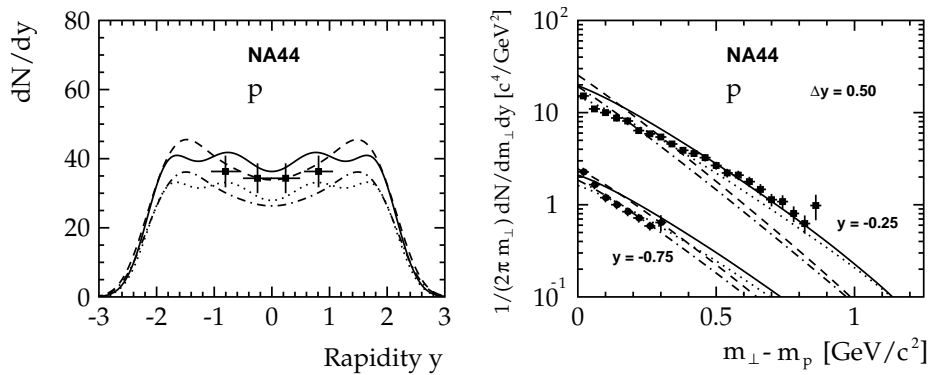}
\vspace*{-1.0cm}
\caption[]{Rapidity spectra, $dN/dy$, and transverse mass
spectra, $1/m_\perp dN/dm_\perp$, for protons, $p$, respectively.
The solid (dotted) lines indicate the results of the calculations when using 
equation of state EOS-I including (without) contributions from $\Lambda^0$ 
decay, whereas the dashed (dashed-dotted) lines represent the results when 
using equation of state EOS-II including (without) contributions from 
$\Lambda^0$ decay. 
The data have been taken by the NA44 Collaboration \cite{NA44baerden}.
}
\vspace*{-0.5cm}
\label{fig06}
\end{figure}

\hspace*{-0.6cm}$v_\perp^{max}(II)$ = 0.30 $c$.

Since the data analysis for the 158 $AGeV$ Pb+Pb collisions has not been
finished yet by the NA44 and NA49 Collaborations, we cannot draw too many 
conclusions here. 
Therefore, in the following we shall look into the space-time 
features of the hydrodynamical model calculations and try to learn, whether 
Bose-Einstein correlations provide a more distinct observable in the attempt to 
decide which one of the two considered EOS gives a better description of the data.

\section{Space-Time Geometry and Bose-Einstein Correlations}
 
In Fig.~7 it can be seen, that the hydrodynamical solutions which are provided 
from the numerical analysis represent an evolution of initally disk-shaped 
fireballs which emit hadrons from the very beginning of their formation. While 
the relativistic fluids expand in longitudinal and in transverse directions,
the longitudinal positions of the freeze-out points increase their distance 
relative to the center. Because of the effect of transverse inwardly moving 
rarefaction waves, the transverse freeze-out positions move towards the center
of the fireball. In the late stage of the hydrodynamic expansion the 
hadron-emitting fireballs separate into two parts while cooling down until
they cease to emit.

For both equations of state we obtain fireballs which have similar transverse
sizes and a QGP phase of similar lifetime. For EOS-I (EOS-II) the total lifetime
of the QGP is $t_{QGP}$ = 2.4 $fm/c$ (3.0 $fm/c$), considering the critical
temperature $T_c$ = 200 $MeV$ (160 $MeV$). However, the softer EOS-II
results in a fireball that has a much larger longitudinal size than the fireball
which is governed by EOS-I.

\begin{figure}
\vspace*{-0.3cm}
\hspace*{-0.3cm}
                 \epsfxsize=13.0cm
		 \epsfbox{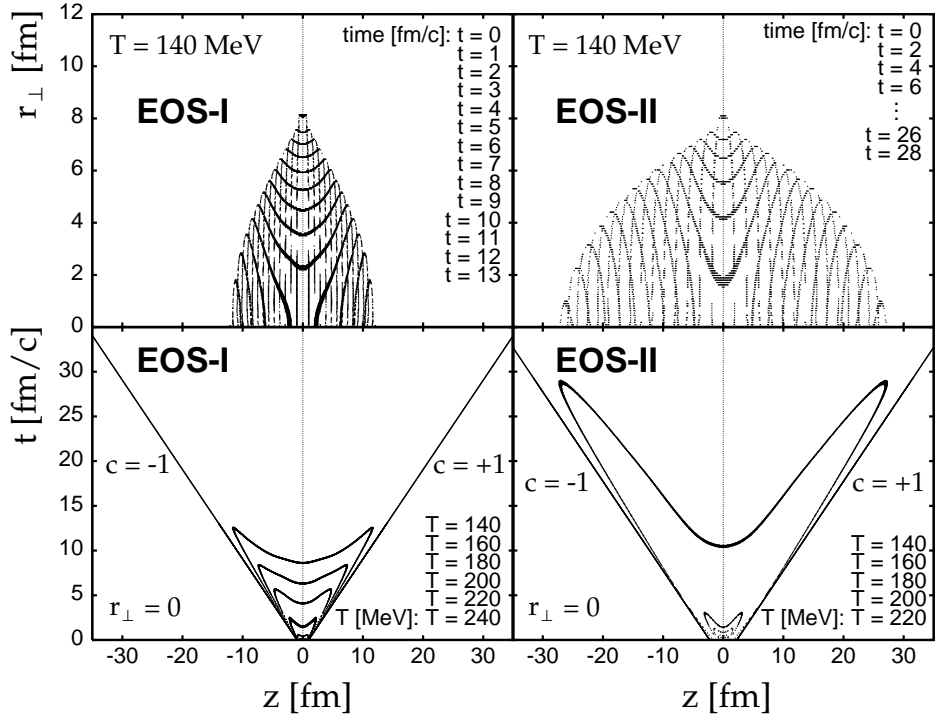}
\vspace*{-0.5cm}
\caption[]{
Time contour plots of the freeze-out hypersurfaces in the $z - r_\perp$
plane, and temperature evolution plots at $r_\perp = 0$, for the relativistic
fluids governed by EOS-I and EOS-II, respectively. In the time contour plots,
each line represents an isotherme ($T_f$ = 140 $MeV$) at a fixed time $t$
(timesteps $\Delta t$ = 1.0 $fm/c$). In the temperature evolution plots, 
each line corresponds to a fixed temperature, $T$ (temperature steps $\Delta T$
= 20 $MeV$).
}
\vspace*{-3.5cm}
\label{fig07}
\end{figure}
\vspace*{2.7cm}The total lifetime of the fireball is $t_{max}$ = 29.3 $fm/c$ (13.1 $fm/c$) 
for EOS-II (EOS-I). Such large differences (factor $\sim$ 2) in the lifetimes and 
longitudinal sizes of the two considered scenarios should show up in the inverse widths 
of the transverse ``out'' \cite{bertsch} and longitudinal Bose-Einstein correlations of 
identical pion pairs. 

In Fig.~8 inverse widths of BEC functions of identical negative pion pairs (including 
decay contributions from resonances) are shown, which have been extracted with a 
Gaussian fit as explained in refs. \cite{bernd14},\cite{chapman}, in comparison
with preliminary measurements of the NA49 Collaboration \cite{NA49kadija}.
The inverse widths are plotted as functions of the average transverse momentum of the 
pion pair, $K_\perp$. The effective longitudinal radii, $R_{long}$, are evaluated in 
the longitudinal comoving system (LCMS), whereas all other effective transverse radii, 
$R_{side}$ and $R_{out}$, have been calculated in the nucleus-nucleus center of mass 
system. Since BEC of negative pions are in general not of Gaussian shape, ({\it cf.},
e.g., refs. \cite{bernd7},\cite{bernd3},\cite{bernd11}), the fit to a Gaussian generates
an error for the inverse width parameters. 

\begin{figure}
\vspace*{-0.3cm}
\hspace*{0.7cm}
                 \epsfxsize=11.0cm
		 \epsfbox{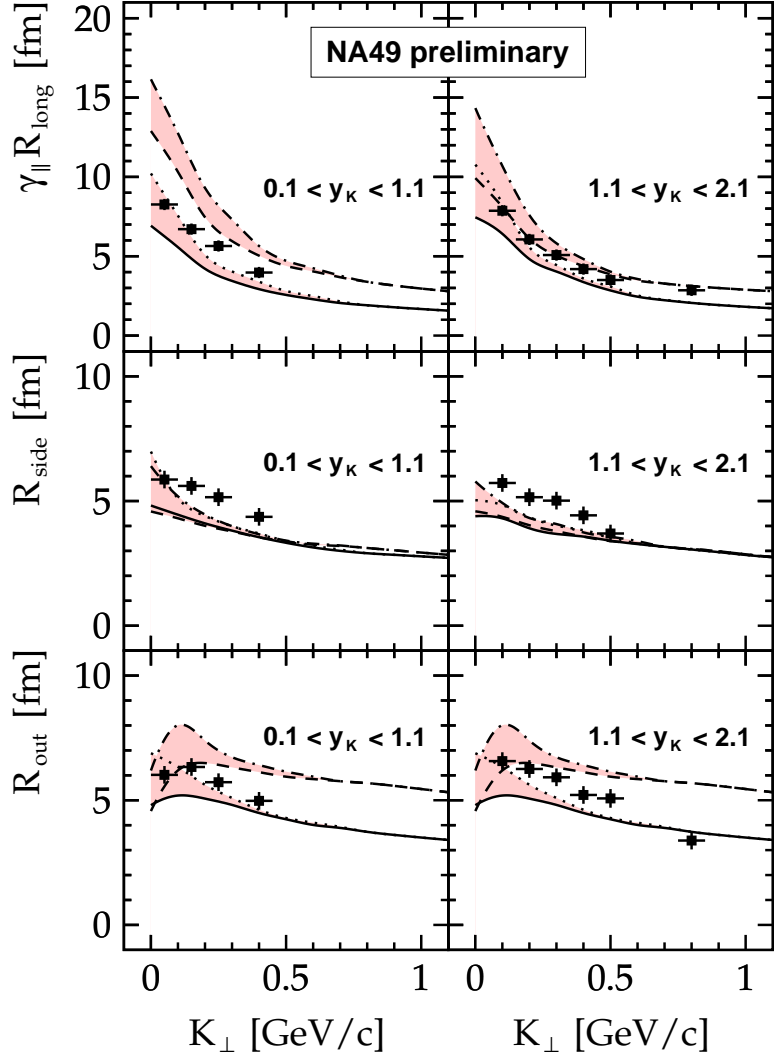}
\vspace*{-0.5cm}
\caption[]{
Inverse widths of BEC functions of identical negative pion pairs (including
decay contributions from resonances) as functions of the average transverse
momentum of the boson pair, $K_\perp$, in the indicated ranges of the
particle pair rapidities, $y_K$, compared to preliminary data of the
NA49 Collaboration \cite{NA49kadija}, respectively. The solid (dashed) lines 
indicate the inverse widths of BEC functions extracted from a Gaussian fit 
for the calculation using EOS-I (EOS-II). The dotted (dashed-dotted) lines
are the true inverse widths of the correlation functions at their 68\% level
(see text), and the grey zones reflect the theoretical uncertainties, when
extracting the inverse widths. 
}
\label{fig08}
\end{figure}

Therefore, in Fig.~8 also the true inverse widths of the correlation functions are 
included at their 68\% level, i.e., the inverse widths at $0.68\cdot(I_0 - 1)$, where 
$I_0 \equiv C_2(\vec{k},\vec{k})$ is the intercept of the BEC function, 
$C_2(\vec{k}_1,\vec{k}_2)$, of two identical pions with momenta $\vec{k}_1$ and
$\vec{k}_2$, respectively.

As expected, the transverse radii $R_{side}$ as functions of $K_\perp$ are very
similar when comparing the calculations for EOS-I with the calculations for EOS-II.
The transverse radii $R_{out}$ and the longitudinal radii $R_{long}$ are
larger for the calculation using EOS-II compared to the calculation using EOS-I.
But the differences, e.g., in the longitudinal effective radii are not so pronounced 
anymore as in the case of the longitudinal sizes of the exhibited space-time geometries 
of the hadron sources. However, they are quite significant. 

Unfortunately, the data analysis of the Bose-Einstein correlations for the 158 $AGeV$ 
Pb+Pb collisions has not been finished yet either by the NA49 Collaboration, so that
the experimental situation does not provide an definite answer yet, which one of
the considered EOS gives a better description of the data.

\section{Conclusions}

The predominantly preliminary single and double inclusive momentum spectra of 158 
$AGeV$ Pb+Pb collisions, recently measured by the NA44 and NA49 Collaborations,
have been reproduced using the relativistic hydrodynamical model HYLANDER-C.
However, the calculations for the much softer EOS-II yield larger slopes in many 
of the hadronic transverse mass spectra compared to the calculations for EOS-I. 
This effect is particularly strong in the transverse mass spectra of protons and 
anti-protons. The initial conditions, which have been found in the numerical 
analysis, differ strongly for the two considered scenarios. The space-time geometries in longitudinal direction and 
in time show a difference in the two calculations of more than a factor of 2. 
In contrast to the common belief \cite{rischke}, the Bose-Einstein correlation 
functions of identical pion pairs do not show such a strong, but still a significant 
sensitivity to the effects of the equations of state.

\section*{Acknowledgement}
I would like to thank the organizers for their invitation to present these results
at the ``Workshop on Hydrodynamics 1997'' at the ECT* in Trento/Italy. Special thanks 
go to D. Strottman and N. Xu for discussions that initiated this work. 
This work has been supported by the U.S. Department of Energy.
 
\section*{Notes} 
\begin{notes}
\item[a]
E-mail: schlei@t2.LANL.gov
\item[b]
HYLANDER-C is an improved version of the original model HYLANDER 
({\it cf.}, refs. \cite{udo},\cite{jan},\cite{web}).
\end{notes}

\vfill\eject

\begin{thebibliography}{99} 

\bibitem{bernd7} B.R. Schlei, U. Ornik, M. Pl\"umer, D. Strottman,
R.M. Weiner, {\it Phys. Lett.} {\bf B376} (1996) 212.

\bibitem{bernd8} U. Ornik, M. Pl\"umer, B.R. Schlei, D. Strottman,
R.M. Weiner, {\it Phys. Rev.} {\bf C54} (1996) 1381.

\bibitem{boal} D.H. Boal, C.-K. Gelbke, B.K. Jennings, {\it Rev. Mod. Phys.}
{\bf 62} (1990) 553.

\bibitem{GGLP} R. Hanbury-Brown and R.Q. Twiss, {\it Philos. Mag.} {\bf 45} (1954) 663;
G. Goldhaber, S. Goldhaber, W. Lee, and A. Pais, {\it Phys. Rev.} {\bf 120} (1960) 300.

\bibitem{NA44xu} Nu Xu for the NA44 Collaboration, {\it Nucl. Phys.} {\bf A610}
(1996) 175c.

\bibitem{NA44baerden} I. G. Baerden et al. (NA44 Collaboration),
``Midrapidity Protons in 158 $AGeV$ Pb+Pb Collisions'', {\it Phys. Lett.} {\bf B},
(in print); CERN Preprint CERN-PPE/96-163.

\bibitem{NA49jones} P.G. Jones and the NA49 Collaboration, {\it Nucl. Phys. }
{\bf A610} (1996) 188c.

\bibitem{NA49kadija} K. Kadija (NA49 Collaboration), {\it Nucl. Phys.} {\bf A610}
(1996) 248c.

\bibitem{clare} R.B. Clare, D.D. Strottman, {\it Phys. Rep.} {\bf 141}
(1986) 177.

\bibitem{udo} U. Ornik, F. Pottag, R.M. Weiner, {\it Phys. Rev. Lett.}
{\bf 63} (1989) 2641. 

\bibitem{jan} J. Bolz, U. Ornik, R.M. Weiner, {\it Phys. Rev.} {\bf C46} (1992) 2047.

\bibitem{web} for additional detail check the web-site
{\tt http://t2.lanl.gov/schlei/hylander.html}.

\bibitem{euler} L.D. Landau, E.M. Lifschitz, ``Fluid mechanics''
(Pergamon, New York, 1959).

\bibitem{fred} F. Cooper, G. Frye, E. Schonberg, {\it Phys. Rev.} {\bf D11} (1975) 192.

\bibitem{bernd3}  J. Bolz, U. Ornik, M. Pl\"umer, B.R. Schlei, R.M. Weiner,
{\it Phys. Rev.} {\bf D47} (1993) 3860.

\bibitem{redlich}  K. Redlich, H. Satz, {\it Phys. Rev.} {\bf D33} (1986)
3747.

\bibitem{phdudo} U. Ornik, Ph.D. thesis, Universit\"at Marburg, 1990.

\bibitem{dipljan} B.R. Schlei, Ph.D. thesis, Universit\"at Marburg, 1994;
{\tt http://t2.lanl.gov/schlei/eprint.html}.

\bibitem{hung} C.M. Hung, E.V. Shuryak, {\it Phys. Rev Lett.} {\bf75} (1995) 4003.

\bibitem{bertsch} G. Bertsch, M. Gong, and M. Tohyama, {\it Phys. Rev.} {\bf C37} (1988) 
1896.

\bibitem{bernd14} B.R. Schlei, D. Strottman, N. Xu, ``The linear correlation 
coefficient vs. the cross term in Bose-Einstein correlations'', Los Alamos 
Preprint LA-UR-97-324, nucl-th/9702011, submitted to {\it Phys. Rev. Lett.} for
publication.

\bibitem{chapman} S. Chapman, P. Scotto, U. Heinz, {\it Phys. Rev. Lett.}
{\bf 74} (1995) 4400; {\it Heavy Ion Phys.} {\bf 1} (1995) 1.

\bibitem{bernd11} B.R. Schlei, {\it Phys. Rev.} {\bf C55} (1997) 954.

\bibitem{rischke} D.H. Rischke, {\it Nucl. Phys.} {\bf A610} (1996) 88c.

\end{thebibliography}
\end{document}